\documentstyle[12pt,aps]{revtex}
\newcommand{\be}{\begin{equation}} 
\newcommand{\en}{\end{equation}}
\newcommand{\bea}{\begin{eqnarray}}
\newcommand{\ena}{\end{eqnarray}}

\newcommand{\hbo}{\hbox to 1 true cm {\hfill } }

\input psfig 
\def\dslash{\partial\kern-.5em\slash}
\def\kslash{k\kern-.5em\slash}
\def\pslash{p\kern-.5em\slash}

\begin{document} 
\vglue 1truecm
  
\vbox{ UNITU-THEP-01/97
\hfill February 4, 1997
}
  
\bigskip
\centerline{\bf Monopoles contra vortices in SU(2) lattice gauge 
theory?\footnote{Supported in part by DFG under contract Re 856/1--3.} 
} 
\bigskip
\centerline{ Kurt Langfeld and Hugo Reinhardt } 
\bigskip
\centerline{ Institut f\"ur Theoretische Physik, Universit\"at 
   T\"ubingen }
\centerline{D--72076 T\"ubingen, Germany.}
\bigskip
  
\begin{abstract}
We show that the scenario of vortex induced confinement of center--projected 
SU(2) lattice gauge theory is not necessarily in conflict with the 
findings in the positive plaquette model. 

\end{abstract}

\bigskip
Recently~\cite{deb96}, it was reported that the vortices of center 
projected SU(2) lattice gauge theory reproduce the full string tension, 
whereas the lattice configurations fail to yield a non-zero string tension, 
if the vortices are suppressed. The authors argue that these vortices 
are the relevant degrees of freedom to confine quarks in the 
fundamental representation. 

In order to be more precise, center--projection was defined 
in~\cite{deb96} on top of Abelian projection. The maximal Abelian 
gauge~\cite{tho76} makes a link variable $U_\mu (x)$ as 
diagonal as possible, and Abelian projection replaces a link variable 
\be 
U_\mu (x) \; = \; \alpha _0(x)  \; + \; i \vec{\alpha } (x) 
\; \hat{n}(x) \vec{\tau } \; , \hbo 
\alpha _0^2 + \vec{\alpha }^2 =1 
\en 
by the Abelian link variable 
\be 
A \; = \; \frac{ \alpha _0(x)  \; + \; i \alpha _3 (x) 
\; \tau ^3 }{ \sqrt{ \alpha _0^2 + \alpha _3^2 } } \; = \; 
\cos \theta (x) \; + \; i \sin \theta (x) \; \tau ^3 \; . 
\en 
Center--projection is then defined by assigning to each link variable 
a value $\pm 1$ according the rule $A (x) \rightarrow \hbox{sign }
(\cos \theta (x) ) $. 
A plaquette is defined to be part of the vortex, if the product of 
the center--projected links is $-1$. 

This result of ref.~\cite{deb96} seems to be in conflict with the 
findings of ref.~\cite{wen89,sta96}, where lattice calculations 
have been performed in the positive plaquette model (PPM) in the 
maximal Abelian gauge. In the PPM, configurations giving rise to plaquettes 
with negative trace are rejected. Nevertheless, this model reproduces 
the linear rising confinement potential. From this observation, 
the authors of ref.~\cite{sta96} concluded that vortices are not 
responsible for the string tension in contradiction to the findings 
of~\cite{deb96}. 

Here we show that the results of the vortex-approach~\cite{deb96} 
and the results of the PPM~\cite{sta96} 
are not necessarily in conflict. For this purpose, let us assume that 
Abelian projection has been performed and consider a 
particular plaquette configuration $P= U_1 U_2 U_3 U_4$, 
which is generated by the Abelian link variables 
\be 
U_{k} \; = \; \cos \phi _k \; + \; i \sin \phi _k \; \tau ^3 \; , 
\hbo k=1 \ldots 4 \; . 
\label{eq:1} 
\en 
The plaquette can be written as 
\be 
P \; = \; \cos \left( \sum _{k=1}^{4} \phi _k \right) \; + \; 
i \sin \left( \sum _{k=1}^{4} \phi _k \right)  \; \tau ^3 \; . 
\en 
It is easy to choose a set of link variables such that 
\be 
\phi _1 \in [0, \frac{\pi}{2} [ \; , \hbo 
\phi _{k=2\ldots 4} \in [\frac{\pi}{2}, \pi [ \; , \hbo 
\sum _{k=1}^4 \phi _k \in [0, \frac{\pi }{2} [ \; \; \hbox{mod} \; \; 
2\pi \; .
\en 
In these cases, the trace of the plaquette $P$ is positive. Hence 
such a configuration contributes in the PPM. 
On the other hand, center projection assigns to the link 
$U_1$ the value $+1$ and to the links $U_{k=2\ldots 4}$ the value 
$-1$ implying that the configuration represents a vortex, since the 
product of the center--projected links yields $-1$. Our example 
shows that the vortex configurations discussed in~\cite{deb96} are not 
excluded in the PPM. This result is obviously not restricted 
to Abelian projected links. 

In order to further clarify the role of the vortices occurring in the 
center--projected lattice theory, we suggest 
to study the center-projection of the positive plaquette model.

\end{document}